\documentclass[]{youtu} 
\PassOptionsToPackage{table}{xcolor}
\usepackage{CJKutf8}
\usepackage{mathpazo}
\usepackage{hyperref}
\usepackage{booktabs}
\usepackage{array}
\usepackage{url}
\usepackage[utf8]{inputenc} 
\usepackage[T1]{fontenc}      
\usepackage{wrapfig}
\usepackage{url}           
\usepackage{amsfonts}       
\usepackage{nicefrac}       
\usepackage{microtype}      
\usepackage[table]{xcolor}
\usepackage{bm}
\usepackage{makecell}
\usepackage{enumitem}
\usepackage{booktabs}
\usepackage{array}
\usepackage{multirow}
\usepackage{bbding}
\usepackage{colortbl}
\definecolor{mygray}{gray}{.92}
\usepackage[ruled,vlined]{algorithm2e}
 
\usepackage{amsmath,amsthm,amssymb}
\usepackage{tcolorbox}
\usepackage{color}
\usepackage{graphicx}
\usepackage{subcaption}
\usepackage{caption}
\usepackage{natbib}
\title{\raisebox{-2.7pt}{\includegraphics[width=1.2em]{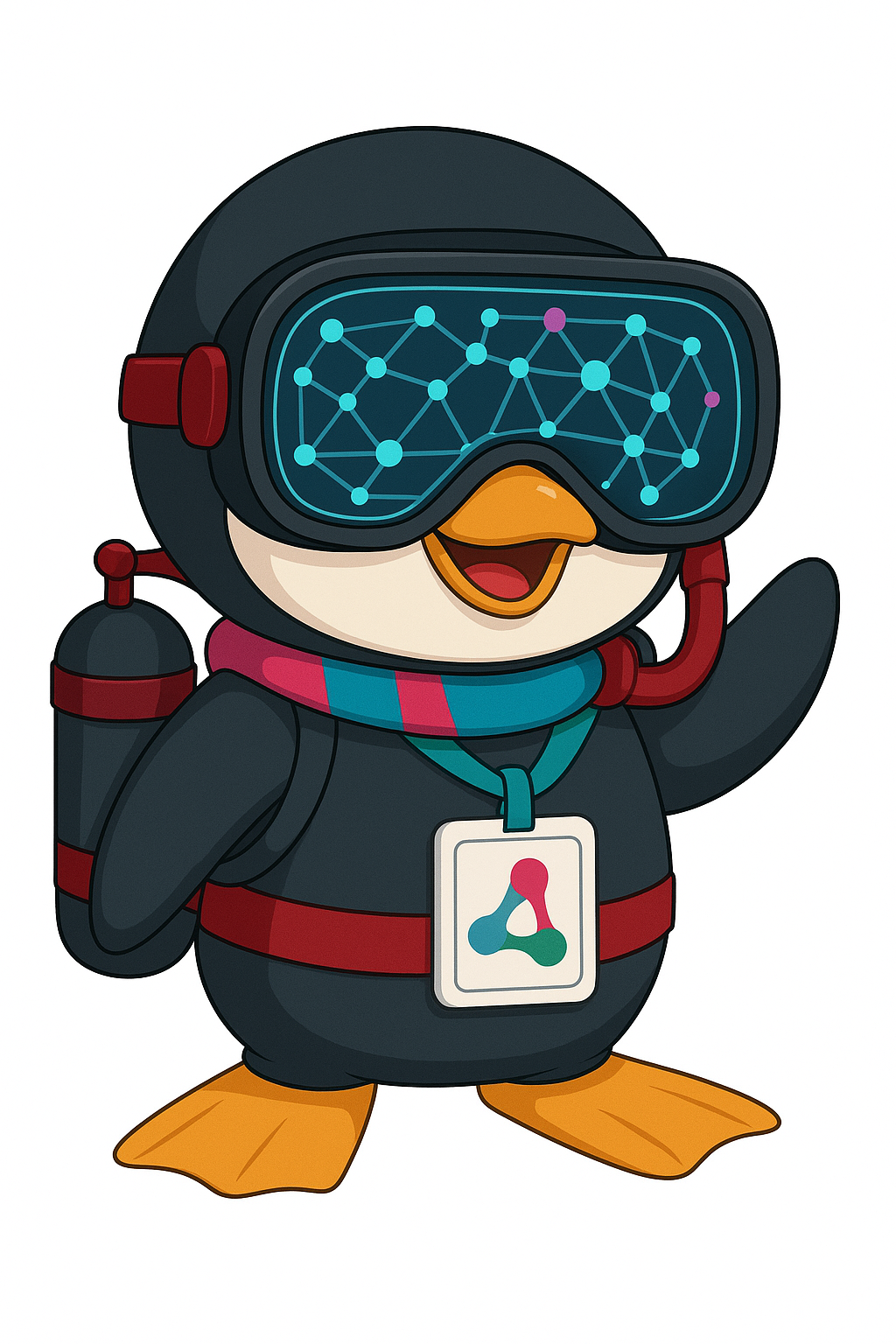}}\xspace \texttt{Youtu-GraphRAG}: Vertically Unified Agents for Graph Retrieval-Augmented Complex Reasoning
}
\author{\small Junnan Dong\textsuperscript{$1 \dagger$}, Siyu An\textsuperscript{$1 \dagger$}\textsuperscript{$\ddagger$}, Yifei Yu\textsuperscript{$1$}, Qian-Wen Zhang\textsuperscript{$1$}, Linhao Luo\textsuperscript{$2$},  \\Xiao Huang\textsuperscript{$3$}, Yunsheng Wu\textsuperscript{$1$}, Di Yin\textsuperscript{$1$},Xing Sun\textsuperscript{$1$}}
\vspace{-5mm}
\affiliation{$^1$Tencent Youtu Lab\\\textsuperscript{$2$}Monash University\\\textsuperscript{$3$}The Hong Kong Polytechnic University}

\sourcecode{https://github.com/TencentCloudADP/Youtu-GraphRAG}
\data{https://huggingface.co/datasets/Youtu-Graph/AnonyRAG}
\youtufinalcopy

\begin{document}

\abstract{
Graph retrieval-augmented generation (GraphRAG) has effectively enhanced large language models in complex reasoning by organizing fragmented knowledge into explicitly structured graphs. Prior efforts have been made to improve either graph construction or graph retrieval in isolation, yielding suboptimal performance, especially when domain shifts occur. In this paper, we propose a vertically unified agentic paradigm, \texttt{Youtu-GraphRAG}, to jointly connect the entire framework as an intricate integration. Specifically, \((i)\) a seed graph schema is introduced to bound the automatic extraction agent with targeted entity types, relations and attribute types, also continuously expanded for scalability over unseen domains; \((ii)\) To obtain higher-level knowledge upon the schema, we develop novel dually-perceived community detection, fusing structural topology with subgraph semantics for comprehensive knowledge organization. This naturally yields a hierarchical knowledge tree that supports both top-down filtering and bottom-up reasoning with community summaries; \((iii)\) An agentic retriever is designed to interpret the same graph schema to transform complex queries into tractable and parallel sub-queries. It iteratively performs reflection for more advanced reasoning; \((iv)\) To alleviate the knowledge leaking problem in pre-trained LLM, we propose a tailored anonymous dataset and a novel `Anonymity Reversion' task that deeply measures the real performance of the GraphRAG frameworks. Extensive experiments across six challenging benchmarks demonstrate the robustness of \texttt{Youtu-GraphRAG}, remarkably moving the Pareto frontier with up to 90.71\% saving of token costs and 16.62\% higher accuracy over state-of-the-art baselines. The results indicate our adaptability, allowing seamless domain transfer with minimal intervention on schema. 
}
\maketitle

\section{Introduction}
\renewcommand{\thefootnote}{}
\footnotetext{$\dagger$ Equal contribution. hansonjdong@tencent.com, siyuan@tencent.com}
\footnotetext{$\ddagger$ Corresponding author.}
\vspace{-5mm}
Graph retrieval-augmented generation (GraphRAG) has emerged as a promising paradigm to enhance large language models (LLMs) with structured knowledge \citep{graphrag-bench,pan2024unifying}, particularly for complex multi-hop reasoning tasks across multiple documents\citep{wang2024knowledge,dong2024knowgpt}. By representing fragmented documents as connected graphs with underlying relations \citep{gretriever,dong2023hierarchy}, GraphRAG enables LLMs to traverse explicit paths among documents and entities, performing complex reasoning that is otherwise infeasible within flat retrieval \citep{graphragsurvey1,graphragsurvey2}. The structured approach effectively addresses critical limitations in conventional RAG (\citep{dong2024modality}), which often struggles with the coherent relations between discrete pieces of information and multi-hop reasoning.

\vspace{-2mm}
The evolution of GraphRAG brings two distinct but equally important trajectories since the foundational work of \citep{graphrag}. First, from the retrieval front, LightRAG \citep{lightrag} pioneered vector sparsification to improve efficiency. While GNN-RAG and GFM-RAG (\citep{mavromatis2024gnn,gfm}) advanced this direction further by incorporating graph neural networks for fine-grained node matching, more recent HippoRAG 1\&2 \citep{hippo,hipporag2} introduced \begin{wrapfigure}{l}{0.38\linewidth} 
\vspace{-5mm}
    \centering
    \includegraphics[width=\linewidth]{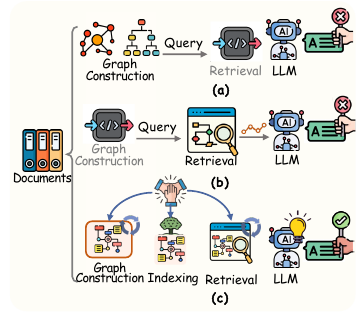}
    \vspace{-5mm}
    \caption{A sketched comparison among existing pipelines and \texttt{Youtu-GraphRAG}. \raisebox{-0.1em}{\includegraphics[height=1em]{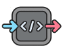}} represents a non-tailored component, indicating current methods focus on either graph construction (a) or retrieval (b) in isolation, while \texttt{Youtu-GraphRAG} proposes a unified paradigm (c) for superior complex reasoning.}
    \label{fig:right-wrap}
    \vspace{-8mm}
\end{wrapfigure} memory and personalized PageRank algorithms for context-aware retrieval. Second, in terms of graph construction, existing methods can be broadly categorized into flat and hierarchical approaches. Early methods, such as KGP~\citep{wang2024knowledge}, rely on existing hyperlinks or KNN-based graphs, resulting in coarse-grained relations that fail to capture nuanced hierarchical semantics. More recent advancements, such as GraphRAG~\citep{graphrag}, combine knowledge graphs with community detection and summarization for multi-level information. Followed by hierarchical methods like RAPTOR~\citep{raptor} and E$^{2}$GraphRAG~\citep{e2}, they further refine the graph using tree-like clustering and recursive summarization to enrich structural representation. However, they remain constrained by their isolated optimizations, concentrating on either construction or retrieval while neglecting their interdependencies. This potentially limits complex reasoning performance where cohesive knowledge organization and retrieval are equally important.

\vspace{-2mm}
To bridge this gap, we aim to answer a critical question:\\
\textbf{\textit{How can we effectively unify graph construction and retrieval for more robust complex reasoning?}}\\ This task is challenging for two reasons. First, construction and retrieval are not readily aligned as two distinct components. It remains difficult to organically establish synergy between them, where the constructed graph could effectively benefit retrieval with both structures and semantics. Second, how to properly evaluate the performance remains a tough problem. With the rapid scaling of LLMs, almost all the existing datasets have already been `seen' before. This fails to reflect the real performance of the entire GraphRAG.
\vspace{-2mm}

In this paper, we propose a vertically unified agentic paradigm, \texttt{Youtu-GraphRAG}, to jointly consider both graph construction and retrieval as an intricate integration based on graph schema. To be specific, \((i)\) a graph schema is introduced to bound the extraction agent that ensures the quality and conciseness with targeted entity types, relations and attribute types; The seed schema is continuously and automatically expanded based on the feedback. \((ii)\) To obtain higher-level knowledge upon the schema, we develop dually-perceived community detection, fusing structural topology with subgraph semantics for comprehensive knowledge clustering. This naturally yields a hierarchical knowledge tree that supports both top-down filtering and bottom-up reasoning with community summaries; \((iii)\) An agentic retriever is designed to interpret the same graph schema to transform complex queries into parallel sub-queries and perform iterative reflection. The agent iteratively performs both reasoning and reflection for more advanced performance; \((iv)\) To alleviate the knowledge leaking problem in pre-trained LLM, we first propose a tailored anonymous dataset with an `Anonymity Reversion' task. Extensive experiments across six challenging benchmarks demonstrate the robustness of \texttt{Youtu-GraphRAG}, remarkably moving the \textit{Pareto frontier} with up to 90.71\% saving of token consumption and 16.62\% higher accuracy over SOTA baselines. The results also indicate our remarkable adaptability which allows seamless domain transfer with minimal intervention on the graph schema, providing insights of the next evolutionary GraphRAG paradigm for real-world applications. 
\vspace{-3mm}

\textbf{Contributions}. In general, our primary contributions are summarized hereunder:\\
\vspace{-8mm}
\begin{itemize}[leftmargin=*]
    \item We first propose a vertically unified Agentic GraphRAG framework to integrate graph construction and retrieval for more robust and advanced reasoning, where both construction and retrieval agents are bounded by graph schema for effective extraction and query decomposition, respectively; \vspace{-2mm}
    \item A novel theoretically-grounded community detection algorithm is employed to inject high-level summarization upon graph schema, simultaneously preserving structural and semantic graph properties; \vspace{-5mm}
    \item We present a tailored anonymous dataset and `Anonymous Revertion' task is proposed to prevent LLM knowledge leaking for fair evaluation of the GraphRAG performance;\vspace{-2mm}
    \item Extensive empirical experiments are conducted over five challenging benchmarks, showing state-of-the-art performance across diverse reasoning tasks and domains that moves the Pareto frontier with up to 90.71\% saving of token costs and 16.62\% higher accuracy.\vspace{-2mm}
\end{itemize}

\begin{figure}
    \centering
    \includegraphics[width=1\linewidth]{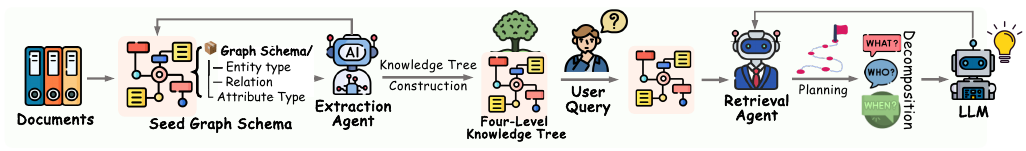}
    \caption{A toy overview of \texttt{Youtu-GraphRAG} that unifies graph construction and retrieval through a schema-guided agentic paradigm. \((i)\) An extraction agent automatically processes documents into structured knowledge via targeted entity/relation extraction; \((ii)\) A four-level knowledge tree is constructed upon the schema with a community detection that fuses topological structures and graph semantics, enabling hierarchical reasoning; \((iii)\) A retrieval agent decomposes user queries into parallel sub-queries aligned with the schema, iteratively driving multi-route retrieval.}
    \label{fig:main}
\end{figure}

\section{Task Definition}
In this section, we formally define the general GraphRAG pipeline with standardized notations from scratch, including both graph construction and graph retrieval. We denote scalars as lowercase alphabets (e.g., $a$), vectors as boldface lowercase alphabets (e.g., $\textbf{a}$), matrices as boldface uppercase alphabets (e.g., ${\bf A}$) and copperplate for a set of elements (e.g., ${\mathcal{A}}$). We refer to GraphRAG as the task of answering a natural language question by first retrieving structured knowledge from a corpus and then generating a response.

\fbox{\parbox{1\textwidth}{
Given a set of documents $\mathcal{D}$, GraphRAG first leverages a frozen LLM $f_{\text{LLM}}(\cdot)$ to extract important knowledge, connected by a structured graph $\mathcal{G}$ as output. To enrich the understanding of $\mathcal{G}$, a community detection algorithm $f_{\text{comm}}(\mathcal{G})$ is employed to partition $\mathcal{G}$ into communities $\mathcal{C}=\{\mathcal{C}_1, \mathcal{C}_2\dots\mathcal{C}_m\}$ to obtain higher-level summarizations. Based on the constructed graph $\mathcal{G}$, given a complex query $q \in \mathcal{Q}$, a retrieval model $f_{\text{retrieve}}(q, \mathcal{G})={\arg\max} \ \mathcal{P}(\mathcal{G}_{\text{sub}} \mid \mathbf{q})$ traverses the graph and retrieves top-$k$ question-specific subgraphs $\mathcal{G}_{sub}\subseteq \mathcal{G}$ that maximize the similarity with given query $q$. The final performance is evaluated from multiple aspects: \((i)\) graph construction costs including time efficiency and token consumptions; \((ii)\) retrieval accuracy and efficiency; and \((iii)\) final answer accuracy comparing $a_{\text{pred}}$ and ground-truths $a_{\text{gold}}$.
}}
\vspace{-5mm}
\subsection{Construction Stage}
\vspace{-5mm}
Beginning with the documents $\mathcal{D}$ as corpus, contemporary GraphRAG research includes two synergistic knowledge organizations that form the graph $\mathcal{G}$ at different granularities. First, the fine-grained graph $\mathcal{G}_\text{triple} = (\mathcal{E}, \mathcal{R}, \mathcal{D})$ is constructed by using $f_{\text{LLM}}(\mathcal{D})$ to extract atomic units in the form of triples $(h, r, t)$ from each document $d \in \mathcal{D}$, where entities $\{h,t\} \in \mathcal{E}$ and relations $r \in \mathcal{R}$ are explicitly linked to represent the abundant relational information among them. The extraction is performed by the frozen LLM $f_{\text{LLM}}(d)$, which processes raw text to populate $\mathcal{G}_\text{triple}$ with schema-compliant triples. Concurrently in another pipeline of research, a coarse-grained document graph $\mathcal{G}_\text{doc} = (\mathcal{D}, \mathcal{C})$ is built by directly clustering documents to maximally preserve the raw context where the atomic units are documents instead of triples. To obtain higher-level knowledge, a complementary community detection algorithm $f_{\text{comm}}(\mathcal{G})$ is employed. Typically, $f_{\text{comm}}(\mathcal{G})$, including Louvain, Leiden, GMM \citep{traag2019louvain,raptor}, etc., operates over $\mathcal{G}$ with sufficient summaries and abstracts generated by $f_{\text{LLM}}(d)$, and results in communities $\mathcal{C}=\{\mathcal{C}_1, \mathcal{C}_2\dots\mathcal{C}_m\}$. $\mathcal{C}_i \subseteq \mathcal{G}$ is further summarized into a high-level meta-node $\hat{e}_i = f_{\text{LLM}}(\mathcal{C}_i)$ by $f_{\text{LLM}}(\mathcal{C})$ where $\hat{e}_i\in \mathcal{G}$. The performance is evaluated by the time used $t_{\text{construct}}$ and the token consumptions $\$$ during construction.
\vspace{-5mm}
\subsection{Retrieval Stage}
\vspace{-5mm}
During inference, given a query $q \in \mathcal{Q}$, the typical retrieval model $f_{\text{retrieve}}(q, \mathcal{G})={\arg\max} \ \mathcal{P}(d \mid {\bf q})$ directly returns the top-$k$ similar documents $\hat{\mathcal{D}}=\{d_1, d_2\cdots d_k\}$ as the final answer, while graph-based methods provide a more explainable subgraph $\hat{\mathcal{G}}$ for multi-hop path traversal, i.e., $f_{\text{retrieve}}(q, \mathcal{G})={\arg\max} \ \mathcal{P}(\hat{\mathcal{G}} \mid {\bf q})$ where $\hat{\mathcal{G}} = \{e_0 \xrightarrow{r_1} e_1 \xrightarrow{r_2} \cdots \xrightarrow{r_k}e_k\} \in \mathcal{G}$. Based on the retrieved subgraph, $f_{\text{LLM}}(q,\hat{\mathcal{G}})$ is employed to generate the final answer. The final performance is evaluated holistically by the retrieval recall comparing $\hat{\mathcal{G_\text{doc}}}$ and ground truth documents $\mathcal{A}^{\text{doc}}_{\text{gold}}$ and answer accuracy by comparing between $a_\text{pred}$ and $a_\text{gold}$.

\section{Approach: \texttt{Youtu-GraphRAG}}
\vspace{-1mm}
In this section, we elaborate on the core methodology of \texttt{Youtu-GraphRAG}, designed to answer two fundamental research questions: \((i)\) How to achieve unified optimization of graph construction and retrieval for higher robustness and generalizability? \((ii)\) How could we enable effective reasoning across different knowledge granularities?  Correspondingly, our framework integrates three designs in a vertically unified manner based on \textbf{graph schema}. First, a graph schema-bounded agent is designed to ensure construction quality while eliminating noise through automatic expansion. Second, beyond the schema, we present a dual-perception community detection that jointly analyzes both topological and semantic similarity to create multi-scale knowledge clusters which forms a four-level knowledge tree. Finally, an agentic retriever is designed to effectively decompose questions into schema-aligned atomic sub-queries with parallel retrieval routes and iterative reflection.

\subsection{Schema-Bounded Agentic Extraction}
Existing GraphRAG methods leverage either pure LLMs or OpenIE (\citep{hippo,hipporag2,gfm,graphrag}) for named entity recognition and triple extraction. However, this open-ended approach would inevitably introduce noise and irrelevant trivia, thereby reducing the usability of the graph. Instead, we treat graph extraction as constrained generation based on a high-quality seed graph schema for domain-specific tasks and define a compact schema as
\begin{equation}
\mathcal{S}\triangleq\bigl\langle\mathcal{S}_{e},\mathcal{S}_{r},\mathcal{S}_{\text{attr}}\bigr\rangle,
\end{equation}
where $\mathcal{S}_{e}$ indicates the targeted entity types (e.g., \texttt{Person}, \texttt{Disease}), $\mathcal{S}_{r}$ guides the extraction with condensed relations (e.g., \texttt{treats}, \texttt{causes}), and $\mathcal{S}_{attr}$ lists attribute types that could be attached and used to describe any corresponding entities (e.g., \texttt{occupation}, \texttt{gender}). A frozen LLM-based agent $f_\text{LLM}(\mathcal{S},\mathcal{D})$ is bounded to identify matched information that appear in $\mathcal{S}$, effectively reducing the search space to the Cartesian product $\mathcal{S}_{e}\times\mathcal{S}_{r}\times\mathcal{S}_{a}$.  
Formally, for each document $d$, we obtain a set of triples hereunder
\begin{equation}
\mathcal{T}(d)=\bigl\{\,(h,r,t),(e,r_{\text{attr}},e_{\text{attr}})\mid \{f(h),f(t),f(e)\}\in\mathcal{S}_{e},\; \{r,r_{\text{attr}}\}\in\mathcal{S}_{r},\; e_{\text{attr}}\in\mathcal{S}_{\text{attr}}\bigr\}.
\end{equation}
Therefore, in our paper, we define $\mathcal{G}_\text{triple} = (\mathcal{E}, \mathcal{R}, \mathcal{D})$, where the entire entity set $\mathcal{E}=\{\mathcal{E}_r,\mathcal{E}_\text{attr}\}$ contains not only named entities $e$ but also the corresponding attributes $e_{\text{attr}}$ and the relation set $\mathcal{R}$ similarly contains both entity-entity relations $r$ and $r_\text{attr}$, i.e.,  \texttt{has\_attribute} relations to connect entities and attributes. However, a seed schema could be general and require manual efforts for predefinitions, which limits the scalability and adaptability of GraphRAG to unseen domains. We thereby equip the agent with an adding tool and incorporate an adaptive design that dynamically refines the initial schema $\mathcal{S}$ through continuous interaction with the document content. The agent automatically proposes schema expansions by analyzing the underlying relational patterns in each document $d \in \mathcal{D}$ through the update function:
\begin{equation}
    \Delta\mathcal{S} = \langle\Delta\mathcal{S}_e, \Delta\mathcal{S}_r, \Delta\mathcal{S}_{\text{attr}}\rangle = \mathbb{I}[ f_\text{LLM}(d, \mathcal{S}) \odot \mathcal{S}]\geq \mu,
\end{equation}
where \(\mathcal{S}^{(t)}\) represents the schema at iteration $t$, $\mu$ serves as a confidence threshold to control the acceptance of new schema elements. $\Delta\mathcal{S}$ contains candidate expansions for entity types, relations, and attributes, respectively. This dynamic adaptation enables the schema to evolve beyond its initial definitions while maintaining controlled growth, as the agent selectively incorporates only high-confidence patterns that demonstrate sufficient frequency and contextual consistency across documents in the new domain. Therefore, we expect the resulting schema to maintain its compact representation while gaining document-specific expressiveness, effectively balancing between strict schema guidance and flexible knowledge acquisition. Through this mechanism, our framework achieves more comprehensive knowledge coverage compared to static schema approaches, particularly when processing domains with emerging relational patterns.
\begin{figure*}
    \centering
    \includegraphics[width=0.9\linewidth]{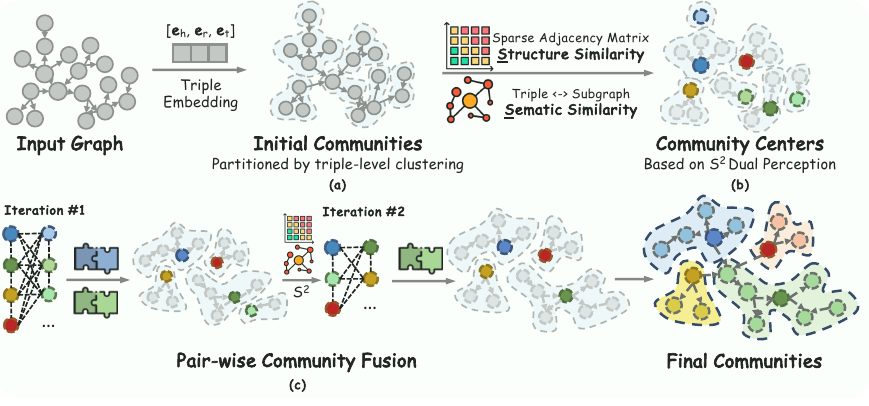}
    \vspace{-3mm}
    \caption{An overview of our dually-perceived community detection. (a) Input graph partitioning into initial communities via triple embeddings; (b) community center identification through joint consideration of topology connectivity and subgraph semantic similarity; and (c) iterative pairwise community merging to form the final hierarchy. Distinct colors represent functionally coherent communities.}
    \label{fig:comm}
\end{figure*}
\subsection{Upon Schema: Graph Indexing with Knowledge Tree}
\vspace{-1mm}

The fine-grained raw graphs could quickly become extremely dense and noisy. Typically, a complementary community detection algorithm $f_{\text{comm}}(\mathcal{G})$ is employed to summarize the knowledge so as to reorganize the graph in communities $\mathcal{C}=\{\mathcal{C}_1, \mathcal{C}_2\dots\mathcal{C}_m\}$. Contemporary methods apply Louvain, Leiden, Gaussian Mixture Models (GMM) (\citep{traag2019louvain,raptor}), etc., operates over $\mathcal{G}$ with sufficient summaries and abstracts generated by $f_{\text{LLM}}(d)$. $\mathcal{C}_i \subseteq \mathcal{G}$ is further summarized into a high-level meta-node $\hat{e}_i = f_{\text{LLM}}(\mathcal{C}_i)$ by $f_{\text{LLM}}(\mathcal{C})$ where $\hat{e}_i\in \mathcal{G}$. 

However, the performance of existing community detection methods can hardly satisfy the real-world demands. They primarily rely on structural connectivity while largely neglecting the rich semantic information embedded in the relational context. As a result, they often produce suboptimal partitions in real-world knowledge graphs since both topological and semantic coherence are crucial for meaningful community detection. To address this limitation, we are motivated to propose a novel and revolutionary dual-perception community detection framework that simultaneously optimizes for topological connectivity and semantic similarity through a three-stage optimization process. An illustration is shown in Figure~\ref{fig:comm}. The final output is compressed into a \emph{Knowledge Tree} $\mathcal{K}$ of depth $L$ that preserves fine-grained facts at the leaves and coarse summaries at internal nodes. In our paper, we define $L=4$, including $\{$\texttt{Community, Keywords, Entity-relation Triples, Attributes}$\}$.

\textbf{Entity Representation}.  Given a graph $\mathcal{G} = (\mathcal{E}, \mathcal{R})$, we first encode each entity $e_i \in \mathcal{E}$ by harvesting its contextualized embedding $\mathbf{e}_i \in \mathbb{R}^{3d}$, aggregating the frozen LLM embeddings of all triples within its one-hop neighborhood $\mathcal{N}_i$. Specifically, for each triple $(e_i, r, e_j) \in \mathcal{N}_i$, we concatenate the embeddings of the head entity $\mathbf{e}_i$, relation $\mathbf{r}_{ij}$, and tail entity $\mathbf{e}_j$, then average across all neighboring triples:   
\begin{equation}
\mathbf{e}_i = \frac{1}{|\mathcal{N}_i|} \sum_{(e_i,r,e_j) \in \mathcal{N}_i} \bigl[ \mathbf{e}_i \| \mathbf{r}_{ij} \| \mathbf{e}_j \bigr].
\end{equation}
To this end, the entity representation could effectively preserve both local structural patterns and semantic relations, enabling downstream clustering to leverage both signals.  

\textbf{Cluster Initialization}. Due to the huge size of real-world graph $\mathcal{G}$, we first reduce the search space by initializing the communities by applying K-means clustering on the entity embeddings $\{\mathbf{e}_i\}_{i=1}^N$, producing an initial partition candidates $\{\mathcal{C}_1^{(0)}, \dots, \mathcal{C}_k^{(0)}\}$, where the superscript denotes the iteration count. While this step provides a coarse grouping, it does not yet account for the interplay between structural and semantic similarity. The cluster number is limited as $k = \min\Big(\max\big(2, \lfloor \frac{|\mathcal{E}|}{\beta} \rfloor\big), \eta\Big)$, where $\beta$=10 controls the granularity that ensures minimum 10 entities per cluster, $\eta$=200 prevents excessive fragmentation. We implement this with optimized KMeans (\texttt{n\_init=5}, \texttt{random\_state}=42) to ensure reproducibility.

\textbf{Iterative Community Fusion via Dual-Perception Scoring}. First, to refine the initial clusters, we introduce a dual-perception scoring function $\phi(e_i, \mathcal{C}_m^{(t)})$ that quantifies the affinity between a node $e_i$ and a community $\mathcal{C}_m^{(t)}$ at iteration $t$. This score combines two considerations. \((i)\) \textbf{topological connectivity overlap} ($\mathbb{S}_r$) that measures the Jaccard similarity between the relation incident to $e_i$ and those in $\mathcal{C}_m^{(t)}$; \((ii)\) \textbf{subgraph semantic similarity} ($\mathbb{S}_s$), which computes the cosine similarity between the entities’s embedding $\mathcal{F}_\Theta(\mathbf{T}_i)$ and the community centroid $\mathbb{E}_{\mathcal{C}_m^{(t)}}[\mathcal{F}_\Theta(\mathbf{T}_{jk})]$, where $\mathcal{F}_\Theta$ is a matrix for embedding transformation.
\begin{equation}
\phi(e_i, \mathcal{C}_m) = \underbrace{\mathbb{S}_r(e_i, \mathcal{C}_m)}_{\text{relational}} \oplus \lambda\,\underbrace{\mathbb{S}_s(e_i, \mathcal{C}_m)}_{\text{semantic}},
\end{equation}
with  
\begin{equation}
\begin{gathered}
\mathbb{S}_r(e_i, \mathcal{C}_m) = \frac{\|\Psi(e_i)\cap\Psi(\mathcal{C}_m)\|_2}{\|\Psi(e_i)\cup\Psi(\mathcal{C}_m)\|_2}, \\
\mathbb{S}_s(e_i, \mathcal{C}_m) = \phi\!\Bigl(\mathcal{F}_{\Theta}(\mathbf{T}_i),\; \sum_{j \in\mathcal{C}_m}\bigl(\mathcal{F}_{\Theta}(\mathbf{T}_{j})\bigr)\Bigr),
\end{gathered}
\end{equation}
where $\mathbb{S}_s$ denotes the Jaccard similarity matrix computed over the multiset of incident relation types $\Psi(\cdot)$. $\mathbb{S}_s(i,j)$ measures the overlap of relation-specific neighborhoods between nodes $i$ and $j$.

Leveraging the dual-perception score, at each iteration $t$, we first locate the most representative centroid entities for each community, which maximizes its dual-perception affinity score $\phi \space(e_i, \mathcal{C}_m)$ with respect to the entire community subgraph. We define the center nodes as:  i.e., $e^*_{\text{center}
} = \mathop{\arg\max} \space \phi(e_i, \mathcal{C}_m)$, where ${e_i \in \mathcal{C}_m}$,$\phi(e_i, \mathcal{C}_m)$ is the dual-perception score as aforementioned, combining both topological relation overlap $\mathbb{S}_r(e_i, \mathcal{C}_m)$ and semantic similarity $\mathbb{S}_s(e_i, \mathcal{C}_m)$. This selection criterion ensures that the center node not only exhibits strong structural connectivity within the community, i.e., high $\mathbb{S}_r$ but also encapsulates the dominant semantic characteristics of the subgraph, i.e., high $\mathbb{S}_s$. The resulting center nodes are then employed to serve as high-quality representatives for their respective communities, facilitating efficient pair-wise community fusion. We then facilitate the pairwise matching between all clusters using their centroid dual-perception score. Clusters $(\mathcal{C}_a^{(t)}, \mathcal{C}_b^{(t)})$ are merged if their dual-perception divergence falls below a threshold $\epsilon$:  
\begin{equation}  
\mathbb{E}[\phi(e_i, \mathcal{C}_a^{(t)})] - \mathbb{E}[\phi(e_i, \mathcal{C}_b^{(t)})] < \epsilon.  
\end{equation}  
This design further shrinks the search space from node-community comparison to node-node comparison, yielding a boosted efficient community detection.

\vspace{-1mm}
\subsubsection{Knowledge Tree}
\vspace{-1mm}
To this end, building upon our schema-bounded extraction framework, we develop a hierarchical knowledge organization pipeline that transforms raw graphs into a structured \emph{Knowledge Tree} $\mathcal{K}$. First, the process begins with our novel dual-perception community detection algorithm, which computes entity-community affinity through the combined metric, blending topological connectivity overlap with semantic subgraph similarity. Second, $f_{\text{LLM}}(\mathcal{C}_m)$ is then applied to generate a brief name and description for the entire community based on the member names. These community names are treated as community nodes and inserted into the original graph, connecting with each member entity with the relation \texttt{member\_of}. Third, within each detected community $\mathcal{C}_m$, we identify pivotal keywords by selecting entities maximizing the structural-semantic score $\arg \max_{e_i \in \mathcal{C}_m} \phi(e_i, \mathcal{C}_m)$. 

The resulting hierarchy, together with the schema, collectively informs the construction of our four-layer knowledge tree $\mathcal{K}$. The tree maximizes bottom-up semantic coherence at each level, simultaneously preserving fine-grained reasoning through granular entity-relation/entity-attribute retrieval ($\mathcal{L}_1$) and enhancing high-level community-based filtering ($\mathcal{L}_4$). We formally define it as $\mathcal{K} = \bigcup_{\ell=1}^4 L_\ell$
\begin{figure}
    \centering
    \includegraphics[width=0.8\linewidth]{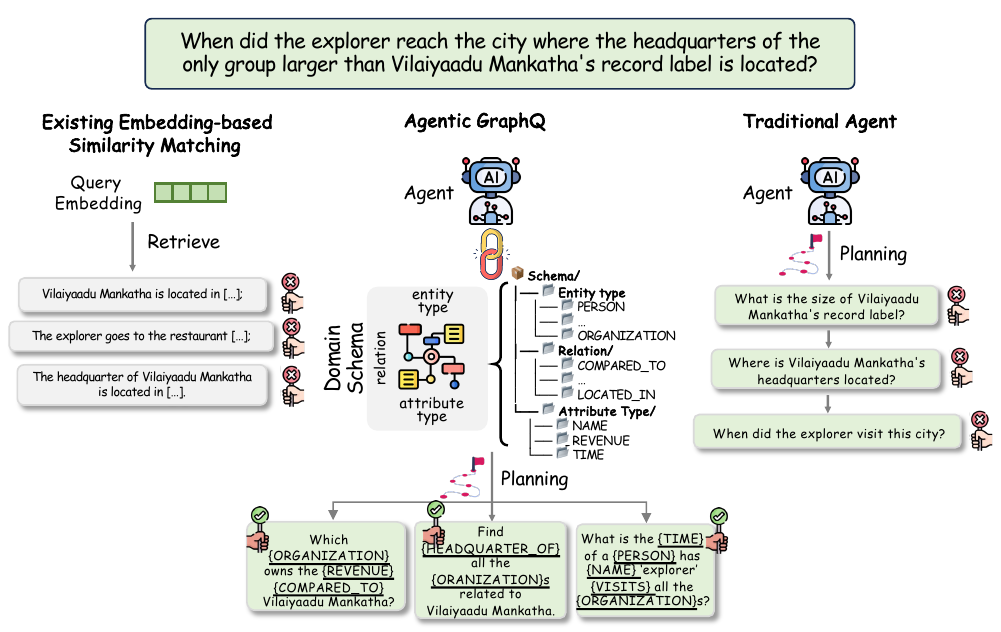}
    \caption{The figure contrasts three query-resolution strategies for a multi-hop question. While embedding matching retrieves disjointed facts (left) and traditional agents use repetitive templates (right), our agentic decomposer (center) leverages domain schema to plan efficient sub-queries: (1) compare record label revenues, (2) locate the larger group’s headquarters, and (3) trace the explorer’s visit—achieving precise, with parallel reasoning and outperforming unstructured retrieval and template-based agents.}
    \label{fig:agent}
\end{figure}
\begin{equation}
L_\ell = \begin{cases}
\{ \mathcal{C}_m \} & \ell=4 \text{ (\texttt{Community})}\\
\{ \arg\max \phi(v_i,\mathcal{C}_m) \} & \ell=3 \text{ (\texttt{Keywords})} \\
\{ (h,r,t) \mid h,t \in \mathcal{E},r \in \mathcal{R} \} & \ell=2 \text{ (\texttt{Entity-Relation Triples})} \\
\{ (e,\texttt{has\_attr},\{e_\text{attr}^{\text{type}}:e_\text{attr}^{\text{value}}\}) \} & \ell=1 \text{ (\texttt{Attributes})}
\end{cases}
\end{equation}

\subsection{Agentic Retriever}
\textbf{Schema-enhanced Query Decomposer}. The complexity of multi-hop queries in large-scale knowledge graphs necessitates an intelligent decomposition mechanism that respects both the explicit schema constraints and implicit semantic relationships. Our schema-guided decomposition approach provides several key advantages over traditional methods. First, by leveraging the graph schema $\mathcal{S} = (\mathcal{S}_e, \mathcal{S}_r, \mathcal{S}_{\text{attr}})$, where $\mathcal{S}_e$ denotes entity types, $\mathcal{S}_r$ represents relation types, and $\mathcal{S}_{attr}$ contains attribute definitions, we ensure that each generated atomic sub-query strictly adheres to valid patterns in the knowledge graph. This schema-awareness prevents the generation of ill-formed queries that would either fail to return results or retrieve irrelevant information. For instance, when processing a query like "Which pharmaceutical companies manufacture diabetes drugs?", the schema guarantees that the "manufacture" relation only connects companies to drugs, not to other entity types. Second, the schema serves as a semantic framework that maintains coherence throughout the decomposition process. Consider the query "Where did Turing Award winners study?" Our method automatically maps "Turing Award winner" to the appropriate entity type $\mathcal{S}_e: \texttt{Person}$ with the specific award attribute, while correctly interpreting "study" as an $\mathcal{S}_r: \texttt{educated\_at}$. This semantic precision prevents the common problem of interpretation drift that often occurs in naive decomposition approaches. Therefore, the final $\mathcal{Q}=f_{\text{LLM}}(q,\mathcal{S})=\{q_1,q_2\dots q_i\}$, where $i$ is a pre-defined maximum number for total atomic sub-queries and each $q_i$ explicitly targets either: \((i)\) node-level retrieval $(e, \text{has\_attr}, a)$, \((ii)\) triple-level matching $(h,r,t)$, or \((iii)\) community-level verification $\mathcal{C}_m$, as determined by schema elements $\mathcal{S}_e, \mathcal{S}_r$, and $\mathcal{S}_{attr}$.  

\textbf{Iterative Reasoning and Reflection}. 
Since reasoning and reflection are two core cognitive capabilities for the agent, following the standard agent framework of perception-reasoning-action cycles, we formalize our agent as a tuple $\mathcal{A} = \langle \mathcal{S}, \mathcal{H}, f_{\text{LLM}} \rangle$, where $\mathcal{H}$ denotes the agent's historical memory containing both reasoning steps and the retrieval results, and the functions $f_{\text{LLM}}$ is employed to implement both key operations. 
\begin{equation}
\mathcal{A}^{(t)} = \underbrace{f_{\text{LLM}}\big( q^{t}}_{\text{Reasoning}},\underbrace{\mathcal{H}^{(t-1)}\big)}_{\text{Reflection}},
\end{equation}
This process addresses the compositional generalization challenge in complex QA by \((i)\) maintaining explicit symbolic grounding through $\mathcal{S}$ during reasoning steps, and \((ii)\) performing continuous self-monitoring via reflection to detect and correct reasoning paths. The agent's operational flow alternates between forward reasoning with schema-guided query decomposition and retrieval and backward reflection for complex scenarios, creating a closed-loop framework that progressively converges to optimal solutions.

\textbf{Multi-Route Retrieval}. To handle diverse sub-query types, we implement four parallel retrieval strategies with distinct optimization objectives:  
\begin{equation}  \small
\begin{aligned}  
\textbf{Entity Matching}: & \quad \arg\max_{e \in \mathcal{E}} \cos\bigl(\mathbf{e}, {\bf q}_i\bigr) \\  
\textbf{Triple Matching}: & \quad \arg\max_{(h,r,t) \in \mathcal{G}} \cos\bigl((\mathbf{e}_h,\mathbf{r}, \mathbf{e}_t), {\bf q_i})\bigr) \\  
\textbf{Community Filtering}: & \quad \arg\max_{\mathcal{C}_m \in \mathcal{K}} \cos \bigl(\mathbf{e}_{\mathcal{C}_m}, {\bf q_i}\bigr) \\  
\textbf{DFS Path Traversal}: & \quad \mathcal{P}(q_i) = e_0 \xrightarrow{r_1} e_1 \xrightarrow{r_2} \cdots \xrightarrow{r_n} e_n \quad \text{s.t.} \quad \forall r_i \in \mathcal{R}, n\leq d  
\end{aligned}  
\end{equation}  
In general, the four retrieval paths exhibit distinct specialization patterns: \((i)\) Entity Matching optimally handles single-hop simple queries requiring precise node identification, e.g., atomic fact check problem; \((ii)\) Triple Matching dominates few-hop reasoning tasks by modeling $(h,r,t)$ compositional semantics, particularly effective for relationship inference; \((iii)\) Community Filtering aims to address global queries, e.g., summarization and cross-domain problems through top-down filtering in the cluster; (4) DFS Path Traversal scales to complex multi-constraint problems, we define the maximum depth $d=5$. This specialization aligns with the cognitive spectrum from atomic facts to complex reasoning scenarios.

\section{Experiments}

\subsection{Evaluation Metrics}
\vspace{-5mm}

Following the workflow of RAG, the evaluation is typically divided into two stages: \((i)\) assess the accuracy of retrieved evidence and \((ii)\) examine the end-to-end performance by evaluating the quality of LLMs responses generated from the retrieved evidence. 
In practical deployment scenarios, where multiple valid retrieval references may exist for identical answers, the latter evaluation paradigm has emerged as the prevailing standard in practical applications. 

Regarding the assessment of LLMs responses, several character-based matching protocols, e.g., recall, EM and F1 score were established. 
To account for semantic deviations caused by minor character variations, where slight textual differences may lead to substantially divergent meanings, we employ DeepSeek-V3-0324 to assess response similarity against ground truth references.

During the reproduction of various GraphRAG frameworks, we observed
experimental results exhibit significant variations depending on the prompts in the LLMs generation stage. Specifically, some frameworks(\cite{e2}) instruct to explicitly reject to answer when retrieved evidence is insufficient, while others(\cite{graphrag-bench,raptor}) allow LLMs to leverage its parametric knowledge or ambiguates the instruction in such cases. Given that most LLMs have been exposed to extensive corpora during pretraining, we identify answering questions based on LLMs' knowledge rather than retrieval mechanism as a critical factor for fairly evaluation - we term \textbf{knowledge leaking}.

To separately assess two critical capabilities:
(1) recognizing knowledge limitations, and (2) leveraging LLMs' parametric knowledge, we therefore implement a dual-mode evaluation on three widely-used datasets:
\vspace{-5mm}
\begin{itemize}[leftmargin=*]
    \item \textbf{Reject mode}. Under this mode, LLMs must reject to answer the question when retrieval fails to provide sufficient evidence. This strictly evaluates the retrieval effectiveness and prevent hallucination.
    \item \textbf{Open mode}. LLMs are allowed to answer using either retrieved content or its inherently parametric knowledge. This maximally measures the overall capability in real-world practical deployment.
\end{itemize}
\vspace{-5mm}
We have reproduced representative baselines and conducted comprehensive evaluations based on the metrics in this work. The corresponding prompts are provided in \autoref{appendix:A}. 
Moreover, the observations further underscore the importance of our proposed AnonyRAG dataset to ensure fair and comprehensive assessment of GraphRAG methods.

\subsection{Datasets}
\vspace{-5mm}

We firstly evaluate \texttt{Youtu-GraphRAG} in dual-mode on three widely used multi-hop QA datasets: HotpotQA (\cite{yang2018hotpotqa}), MuSiQue (\cite{trivedi2022musique}) and 2WikiMultiHopQA (abbreviated as 2Wiki \cite{ho20202wiki}), following the setting in (\cite{hippo,hipporag2}) for fair comparison.

To evaluate the framework's performance across diverse domains, we also employ GraphRAG-Bench(\cite{graphrag-bench}), shorted as G-Bench, a benchmark dataset constructed from textbook corpora. Additionally, to prevent knowledge leaking, we propose two novel bilingual anonymous datasets, i.e., \textbf{AnonyRAG-CHS} and \textbf{AnonyRAG-ENG} and propose a challenging `Anonymous Reversion' task. 

We anonymize specific entity types (e.g., people, locations) in the dataset to break the model’s memory shortcuts and prevent it from relying on pretrained knowledge rather than retrieved evidence. Moreover, we preserve semantic coherence through entity linking, enabling LLMs to maintain discourse comprehension despite anonymized mentions. The construction details of the dataset are documented in \autoref{appendix:B}.
\vspace{-5mm}
\subsection{Baselines}
\vspace{-5mm}

We include three pipelines of research as baselines. \((i)\) Naive RAG, as the standard RAG approach that retrieves top-$k$ document chunks using vector similarity search without any explicit knowledge structuring; \((ii)\) Pure GraphRAG, which builds flat knowledge graphs for retrieval but lacks hierarchical organization, focusing primarily on relational reasoning through graph traversal algorithms, including GraphRAG (\cite{graphrag}), LightRAG (\cite{lightrag}), G-Retriever (\cite{gretriever}) and HippoRAG 1\&2 (\cite{hippo,hipporag2}); \((iii)\) Tree-based GraphRAG, represents hierarchical methods that employ recursive clustering and summarization to construct multi-level knowledge trees including RAPTOR (\cite{raptor}) and E$^2$GraphRAG (\cite{e2}).

To ensure a fair performance comparison, we reproduce
all the baselines and \texttt{Youtu-GraphRAG} with the same setting and evaluate with consistent metrics. In terms of base models, we maintain DeepSeek-V3-0324 and Qwen3-32B as the base LLMs and a lightweight embedding model all-MiniLM-L6-v2.
\vspace{-5mm}

\subsection{Overall Evaluation}
\subsubsection{Comparison of Time and Token Consumption}
For baselines involving graph construction and community detection stages, this section compares their token and time consumption. Unless otherwise specified, all LLM APIs invoked here are based on the DeepSeek-V3-0324 and deployed on identical hardware. All procedures are executed using 32-thread concurrent inference to ensure both the efficiency of graph construction and the fairness of comparisons.

\begin{figure}[htbp]
    \centering
    \begin{subfigure}[b]{0.492\textwidth}
        \includegraphics[width=\textwidth]{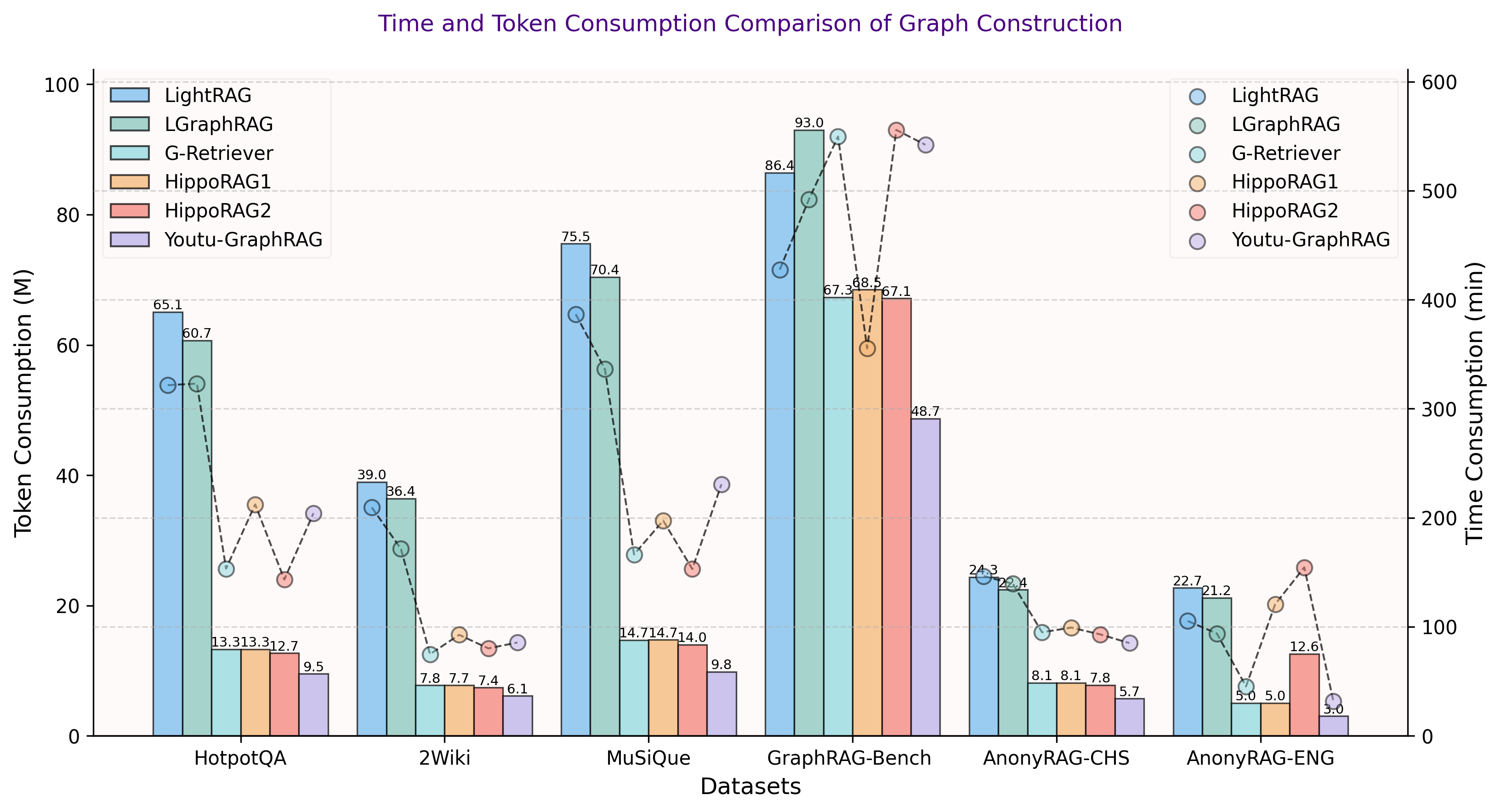}
        \caption{Consumption comparison of graph construction}
        \label{fig:graph_cost}
    \end{subfigure}
    \hfill
    \begin{subfigure}[b]{0.492\textwidth}
        \includegraphics[width=\textwidth]{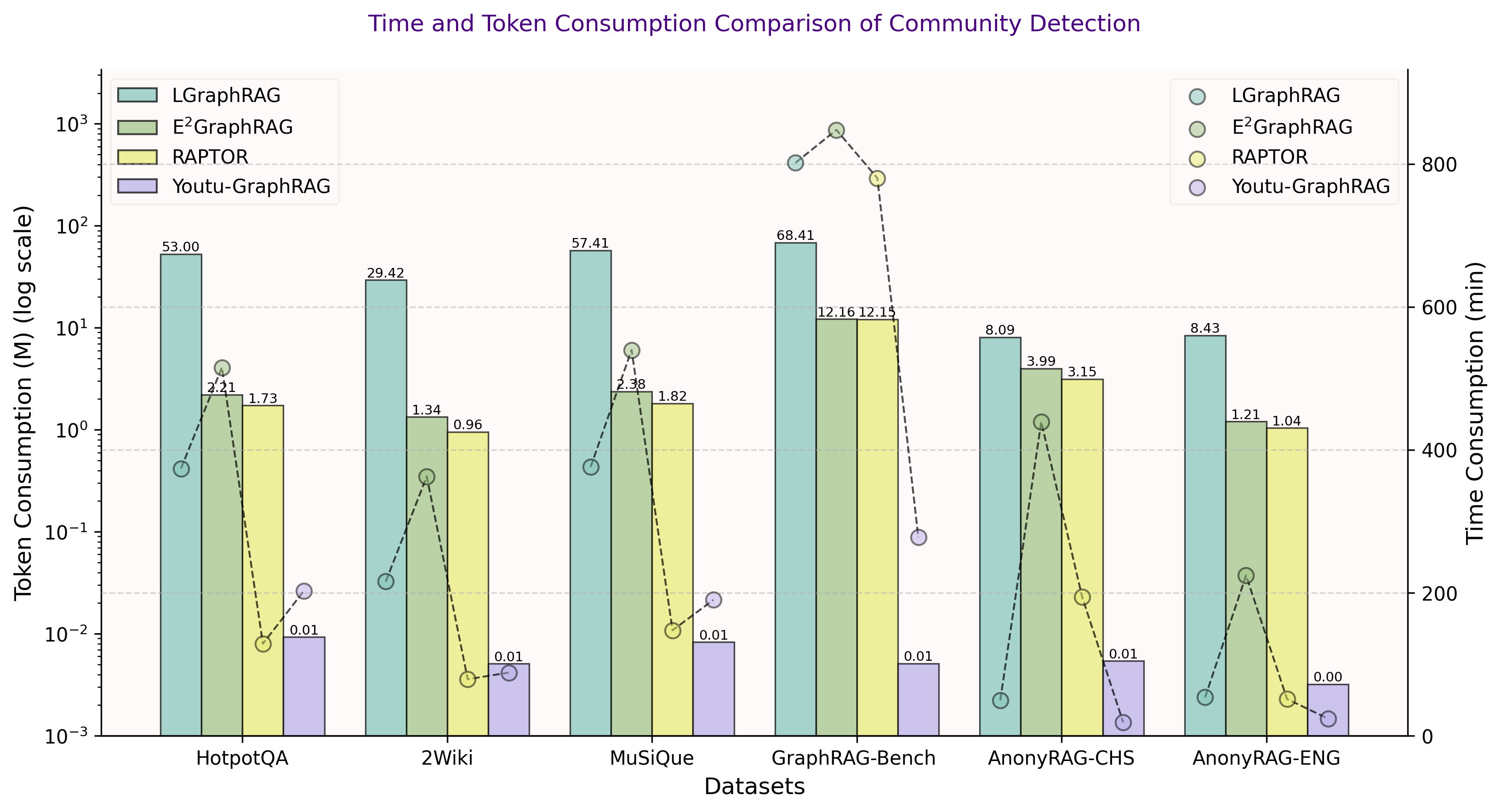}
        \caption{Consumption comparison of community detection}
        \label{fig:community_cost}
    \end{subfigure}
\end{figure}

Figure~\ref{fig:graph_cost} presents the time and token consumption during the graph construction stage for Youtu-GraphRAG and five baselines. Our method consistently achieves the lowest token consumption across all six datasets and maintains relatively efficient time performance on five of the datasets. In the community detection stage, as shown in Figure~\ref{fig:community_cost}, Youtu-GraphRAG achieves the lowest token consumption compared with the other three baselines, consuming no more than 10,000 tokens on any dataset. Meanwhile, our method also demonstrates consistently efficient time performance across all datasets.

\begin{table}[ht!]
\centering
\caption{Overall performance comparisons over benchmark datasets in terms of top-$20$ Accuracy.}
\label{tab:main_results}
\resizebox{0.9\textwidth}{!}{
\footnotesize
\begin{tabular}{l|>{\columncolor{gray!20}}cc|>{\columncolor{gray!20}}cc|>{\columncolor{gray!20}}cc|>{\columncolor{gray!20}}c|c|>{\columncolor{gray!20}}c}
\toprule
\multirow{2}{*}{\textbf{Method}} & \multicolumn{2}{c}{\textbf{HotpotQA}} & \multicolumn{2}{c}{\textbf{2Wiki}} & 
\multicolumn{2}{c}{\textbf{MuSiQue}} & 
\multicolumn{1}{c}{\textbf{G-Bench}} & 
\multicolumn{1}{c}{\textbf{Annoy-CHS}} &
\multicolumn{1}{c}{\textbf{Annoy-ENG}} \\
\cmidrule(lr){2-3} \cmidrule(lr){4-5} \cmidrule(lr){6-7} \cmidrule(lr){8-8} \cmidrule(lr){9-9} \cmidrule(lr){10-10}
& Open & Reject & Open & Reject & Open & Reject & Open & Open & Open \\
\midrule\midrule
\multicolumn{10}{c}{\textbf{Deepseek-V3-0324}}\\
\midrule
Zero-shot LLM & 53.70 & - & 41.6 & - & 25.7 & - & 70.92  & 9.62 & 8.18\\
\midrule
Naive RAG  & 79.90 & 72.40 & 70.3 & 38.9 & 47.49 & 30.63 & 71.81 & 12.5 & 43.02\\
\midrule
E$^{2}$GraphRAG  & 68.70 & 48.80 & 43.20 & 20.00 & 28.36 & 8.01 & 68.66 & 16.01 & 35.97\\
RAPTOR  & 80.90 & 73.60 & 70.10 & 38.40 & 48.50 & 31.10 & 73.08 & 12.08 & 40.2 \\
\midrule
LightRAG & 71.90 & 56.00 & 58.00 & 29.20 & 38.98 & 24.57 & 70.83 & 9.16 & 22.14\\
GraphRAG  & 56.10 & 26.40 & 41.80 & 10.00 & 32.20 & 16.50 & 75.54 & 21.66 & 38.85\\
G-Retriever  & 49.00 & 6.70 & 35.80 & 5.00 & 23.50 & 1.70 & 70.63 & 4.07 & 5.08 \\
HippoRAG  & 81.70 & 73.10 & 77.90 & 64.00 & 48.30 & 36.20 & 72.89 & 36.77 & 40.68\\
HippoRAG-IRCOT  & 81.00 & 74.60 & 78.40 & 66.00 & 46.70 & 35.50 & 73.38 & 36.05 & 42.17\\
HippoRAG2  & 81.80 & 74.90 & 77.30 & 48.30 & 50.80 & 37.80 & 79.37 & 12.92 & 43.16\\
\midrule
\texttt{Ours w/o Agent}  & \textbf{83.70} & \textbf{75.30}& \textbf{72.80} & \textbf{57.80} & \textbf{51.40} & \textbf{40.00} & \textbf{81.53} & \textbf{37.06} & \textbf{40.05}\\
\texttt{Youtu-GraphRAG}  & \textbf{86.50} & \textbf{81.20}& \textbf{85.50} & \textbf{77.60} & \textbf{53.60} & \textbf{47.50} & \textbf{86.54} & \textbf{42.88} & \textbf{43.26}\\
\midrule\midrule
\multicolumn{10}{c}{\textbf{Qwen3-32B}}\\
\midrule
Zero-shot LLM & 36.40  & - & 33.30  & - & 13.40  & - & 70.04 & 5.11 & 6.49\\
\midrule
Naive RAG  & 75.00  & 69.00  & 58.50  & 39.60  & 40.64  & 33.03 & 72.69 & 7.56 & 26.84 \\
\midrule
RAPTOR  & 79.20 & 72.90 & 61.20 & 40.10 & 38.99 & 32.86 & 72.20 & 13.37 & 22.14\\ \midrule
HippoRAG  & 77.00 & 71.80 & 72.80 & 62.50 & 40.60 & 32.10 & 75.64 & 8.58 & 32.30 \\
HippoRAG-IRCOT  & 80.30 & 76.60 & 74.80 & 65.40 & 44.70 & 37.40 & 77.11 & 9.
16& 33.15 \\
HippoRAG2  & 81.80 & 71.30 & 65.20 & 39.90 & 51.40 & 37.70 & 80.35 & 12.65 & 38.36 \\
\midrule
\texttt{Ours w/o Agent}  & \textbf{83.80} & \textbf{73.90}& \textbf{74.90} & \textbf{55.30} & \textbf{52.90} & \textbf{40.10} & \textbf{80.74} & \textbf{34.88} & \underline{35.13}\\
\texttt{Youtu-GraphRAG}  & \textbf{85.90} & \textbf{78.60}& \textbf{85.70} & \textbf{74.20} & \textbf{54.60} & \textbf{45.30} & \textbf{84.48} & \textbf{39.24} & \textbf{40.05}\\
\bottomrule
\end{tabular}}
\end{table}
\subsubsection{Main Performance Comparison}
In Table~\ref{tab:main_results}, we report the top-$20$ accuracy across six challenging benchmarks under both open and reject modes, based on two strong LLM backbones, i.e., DeepSeek-V3-0324 and Qwen3-32b. Across virtually all datasets and settings, \texttt{Youtu-GraphRAG} attains the highest performance, reflecting its ability to combine precise retrieval with robust reasoning. Besides, we also include an variant with no agent for iterative reasoning and reflection as a lightweight version, i.e., \texttt{Ours w/o Agent}, fulfilling real-world applications requiring real-time interactive feedback.

The distinction between the two evaluation modes provides complementary perspectives on system capability. \textbf{Open mode} unlocks the full reasoning potential of the LLM to synthesize an answer regardless of retrieval gaps. This mirrors high-coverage real-world deployments where maximizing end-task accuracy outweighs caution. \texttt{Youtu-GraphRAG} consistently outperforms existing baselines, achieving improvements from 2 to 8 points over the strongest competitor across datasets. When augmented with our agent framework, \texttt{Youtu-GraphRAG} further pushes the performance frontier, reaching top-20 accuracies of 86.5\%, 85.5\%, and 53.6\% on HotpotQA, 2Wiki, and MuSiQue respectively under Deepseek-V3-0324, and 85.9\%, 85.7\%, and 54.6\% under Qwen3-32B, demonstrating a clear advantage in multi-hop reasoning and cross-document synthesis. \textbf{Reject mode}, by contrast, imposes a stringent criterion if the retrieved context is insufficient, the model must abstain. \texttt{Youtu-GraphRAG} attains 81.2\%, 77.6\%, and 47.5\% on HotpotQA, 2Wiki, and MuSiQue, outperforming the strongest baseline by 7–14 points. Across all datasets, our method achieves consistently higher top-20 accuracy, confirming its ability to synergize graph-based retrieval with agent-driven reasoning for both high-coverage and high-precision scenarios. We value this metric since it directly probes retrieval quality, as speculative answers are penalized and the acceptance rate becomes a direct function of retrieval completeness and precision. Our superiority on two anonymous datasets also validates the generalizability of \texttt{Youtu-GraphRAG} beyond standard benchmarks. Specifically, under the open mode, it achieves 42.88\% and 43.26\% top-$20$ accuracy on Annoy-CHS and Annoy-ENG, respectively, surpassing all baselines by a clear margin. These results also reflect our robust reasoning and retrieval integration across diverse languages and domains, demonstrating that our approach could be easily transferred to previously unseen data distributions while maintaining high accuracy. 

\begin{wrapfigure}{l}{0.5\linewidth} 
\vspace{-5mm}
    \centering
    \includegraphics[width=\linewidth]{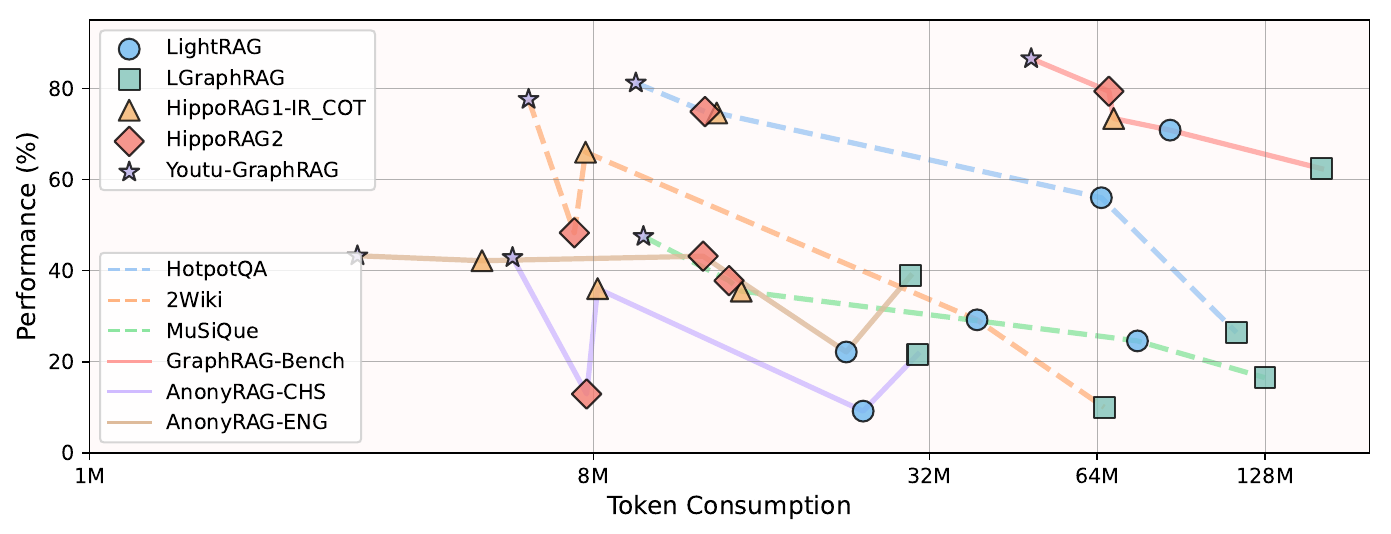}
    \vspace{-5mm}
    \caption{\texttt{Youtu-GraphRAG} effectively moves the Pareto frontier with lower token costs and higher performance.}
    \label{fig:pareto}
    \vspace{-5mm}
\end{wrapfigure}

A key objective of \texttt{Youtu-GraphRAG} is to jointly optimize performance and efficiency by unifying graph construction and retrieval. Figure~\ref{fig:pareto} illustrates the trade-off between token consumption during the construction and overall QA performance across six benchmarks. Our approach consistently achieves optimal performance with the least token consumption, effectively shifting the \textit{Pareto frontier} compared to all baselines.

While existing GraphRAG methods face a dilemma to balance the token consumption during construction and the accuracy for final generation, \texttt{Youtu-GraphRAG} leverages a vertically unified novel framework, i.e., schema-guided extraction, dually-perceived community detection and the schema-enhanced agentic retrieval to build concise yet semantically rich graphs and allow the agent to maximize reasoning effectiveness. As a result, our method effectively moves the \textit{Pareto frontier} and attains the best performance on all benchmarks while consuming up to an order of magnitude fewer tokens during graph construction. This demonstrates that careful integration of structured schema alignment, hierarchical knowledge tree, and adaptive retrieval can fundamentally improve the cost-effectiveness of GraphRAG systems in real-world applications.

\begin{table}[ht!]
\centering
\caption{Overall performance comparisons over benchmark datasets based on DeepSeek in terms of top-$10$ Accuracy.}
\label{tab:top10}
\resizebox{0.9\textwidth}{!}{
\footnotesize
\begin{tabular}{l|>{\columncolor{gray!20}}cc|>{\columncolor{gray!20}}cc|>{\columncolor{gray!20}}cc|>{\columncolor{gray!20}}c|c|>{\columncolor{gray!20}}c}
\toprule
\multirow{2}{*}{\textbf{Method}} & \multicolumn{2}{c}{\textbf{HotpotQA}} & \multicolumn{2}{c}{\textbf{2Wiki}} & 
\multicolumn{2}{c}{\textbf{MuSiQue}} & 
\multicolumn{1}{c}{\textbf{G-Bench}} & 
\multicolumn{1}{c}{\textbf{Annoy-CHS}} &
\multicolumn{1}{c}{\textbf{Annoy-ENG}} \\
\cmidrule(lr){2-3} \cmidrule(lr){4-5} \cmidrule(lr){6-7} \cmidrule(lr){8-8} \cmidrule(lr){9-9} \cmidrule(lr){10-10}
& Open & Reject & Open & Reject & Open & Reject & Open & Open & Open \\
\midrule
Naive RAG  & 79.40 & 68.00 & 67.60 & 33.70 & 45.58 & 26.73 & 71.22 & 12.08 & 38.93\\
\midrule
RAPTOR  & 78.20 & 67.10 & 67.40 & 36.40 & 45.88 & 30.03 & 72.79 & 11.77 & 33.99 \\\midrule
G-Retriever & 49.90 & 5.90 & 38.00 & 3.80 & 23.50 & 1.70 & 70.24 & 5.38 & 5.50\\
LightRAG & 71.98 & 58.10 & 65.70 & 38.10 & 39.40 & 22.90 & 69.74 & 8.58 & 18.90\\
GraphRAG  & 54.30 & 23.70 & 40.00 & 9.80 & 30.20 & 16.00 & 61.39 & 21.37 & 38.36\\
HippoRAG  & 78.20 & 69.40 & 77.10 & 61.10 & 45.20 & 30.90 & 70.14 & 34.01 & 40.12\\
HippoRAG-IRCOT  & 78.10 & 70.20 & 77.70 & 60.70 & 44.40 & 31.60 & 72.89 & 36.19 & 41.42\\
HippoRAG2  & 79.40 & 70.40 & 74.60 & 45.80 & 49.10 & 34.00 & 77.21 & 13.52 & 37.24\\
\midrule
\texttt{Ours w/o Agent}  & \textbf{80.50} & \textbf{72.10}& \textbf{72.10} & \textbf{54.40} & \textbf{49.80} & \textbf{38.30} & \textbf{80.55} & \textbf{35.17} & \textbf{40.54}\\
\texttt{Youtu-GraphRAG}  & \textbf{83.40} & \textbf{78.90}& \textbf{82.30} & \textbf{72.60} & \textbf{52.10} & \textbf{46.90} & \textbf{83.50} & \textbf{38.08} & \textbf{42.57}\\
\bottomrule
\end{tabular}}
\end{table}

\subsection{Analysis of Generalizability}
\begin{wrapfigure}{l}{0.5\linewidth} 
\vspace{-5mm}
    \centering
\includegraphics[width=\linewidth]{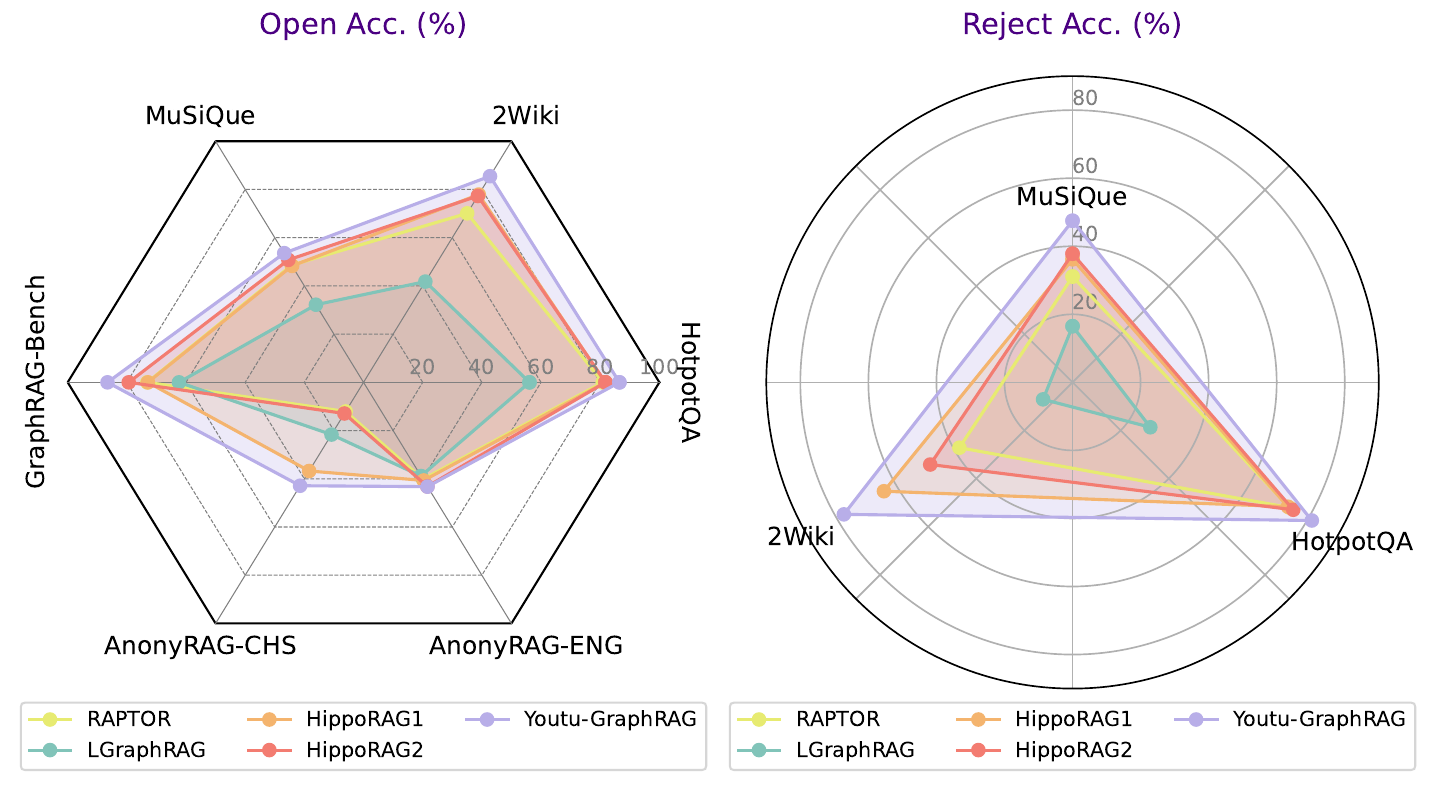}
    \vspace{-5mm}
    \caption{We showcase the generalizability over six benchmark datasets in terms of both open and reject accuracy.}
    \label{fig:radar}
    \vspace{-5mm}
\end{wrapfigure}

To examine the domain-transfer capability of \texttt{Youtu-GraphRAG}, we evaluate it across six heterogeneous benchmarks without any task-specific fine-tuning. As shown in Figure~\ref{fig:radar}, \texttt{Youtu-GraphRAG} achieves the best performance in both Open Accuracy and Reject Accuracy on all datasets, surpassing state-of-the-art GraphRAG baselines by a clear margin.

We attribute this strong generalizability to the intrinsic integration of graph construction and retrieval within our framework. \((i)\) The schema-guided extraction agent produces consistent, domain-adaptive graphs; the dually-perceived community detection yields hierarchical knowledge structures that remain robust across domains; \((ii)\) The agentic query decomposer dynamically adapts retrieval strategies to different question types without manual tuning. Notably, our model demonstrates particularly large gains on multi-hop reasoning datasets such as HotpotQA and 2Wiki in open settings, and shows superior abstention capability on MuSiQue and 2Wiki in reject settings, indicating robustness in both complex reasoning and uncertainty calibration. These results confirm that \texttt{Youtu-GraphRAG} can seamlessly transfer to unseen domains while preserving structural fidelity and reasoning depth, fulfilling the vision of a foundational GraphRAG paradigm.

Furthermore, we summarize the top-$10$ results in Table~\ref{tab:top10}, which provides a more stringent evaluation of retrieval. Our methods consistently outperform all baselines across both open and reject modes. In the open mode, \texttt{Youtu-GraphRAG} achieves top-10 accuracies of 83.4\%, 82.3\%, and 52.1\% on HotpotQA, 2Wiki, and MuSiQue respectively, surpassing the strongest competitor by 4\~8 points. Under the reject mode, the gains are even more pronounced, with improvements of 8\~12 points, indicating robust retrieval fidelity and reduced speculative answering. Notably, on the two anonymous datasets, Annoy-CHS and Annoy-ENG, our agent-enhanced model attains 38.08\% and 42.57\%, reinforcing its consistent superiority in diverse scenarios. These top-$k$ results confirm that our approach not only excels in high-coverage settings but also maintains precise answer selection under stricter evaluation criteria, further validating the effectiveness of integrating graph-based retrieval with agent-guided reasoning.
\begin{table}[t]\footnotesize
\centering
\caption{Ablation studies of our method over six datasets. We evaluate three variants: without Community detection (w/o Comm.), without Agent coordination (w/o Agent), and without Schema guidance (w/o Schema).}
\resizebox{0.9\textwidth}{!}{%
\label{tab:ablation}
\begin{tabular}{lcccccc}
\toprule
\textbf{Variants} & \textbf{HotptQA} & \textbf{2Wiki} & \textbf{MuSiQue} & \textbf{G-Bench} & \textbf{AnonyRAG-CHS} & \textbf{AnonyRAG-ENG} \\
\midrule
w/o Comm. & 79.50 & 75.10 & 44.00 & 85.02 & 39.97 & 39.92\\
w/o Agent & 75.30 & 57.80 & 40.00 & 81.53 & 37.60 & 40.05\\
w/o Schema & 77.10 & 73.40 & 45.60 & 83.50 & 35.61 & 40.32\\
\texttt{Youtu-GraphRAG} & \textbf{81.20} & \textbf{77.60} & \textbf{47.50} & \textbf{86.54} & \textbf{42.88} & \textbf{43.26}\\
\bottomrule
\end{tabular}}
\end{table}
\subsection{Ablation Studies}
To quantify the contribution of each component, we perform ablations by removing community detection (w/o Comm.), agent reasoning and reflection (w/o Agent), and schema guidance (w/o Schema). Results on six benchmarks are summarized in Table~\ref{tab:ablation}.

Specifically, removing community detection leads to a consistent drop across all datasets, particularly on multi-hop QA tasks such as HotpotQA and 2Wiki around 1.7\% and 2.5\%, indicating that structuring knowledge into coherent communities facilitates more accurate retrieval and reasoning for global questions. The absence of agent reasoning and reflection causes the most severe degradation on complex reasoning datasets, especially on 2Wiki and MuSiQue with remarkable 19.8\% and-7.5\% differences, supporting our motivation that the iterative reasoning-feedback loop plays an essential role for resolving ambiguous intermediate steps. Eliminating schema guidance results in noticeable performance drops on knowledge-intensive settings, especially on AnonyRAG-CHS with 7.27\% decreases, highlighting the importance of a high-quality initialization of seed schema for new domains. This further demonstrates our advantage since \texttt{Youtu-GraphRAG} only requires minimum manual intervention to handle with domain shifts. In conclusion, our model consistently outperforms all ablated variants, demonstrating that the three components are complementary: community detection improves retrieval quality, agent reasoning enhances multi-step inference, and schema guidance enforces structural fidelity. These findings suggest that removing any single component disrupts the synergy between retrieval and reasoning, with agent reasoning being most critical for multi-hop inference, while schema plays a vito role in low-resource and domain-specific scenarios.

\section{Related Work}

While large language models (LLMs) demonstrate remarkable capabilities in language understanding and reasoning, they are known to be prone to hallucinations—generating confident yet factually incorrect outputs—especially when reasoning over complex or multi-hop queries~\citep{zhang2025survey,DBLP:journals/corr/abs-2404-04925,DBLP:journals/csur/KuangSXLXLLCLH25, dong2024clr,DBLP:journals/corr/abs-2405-12819}. Integrating LLMs with graph-structured knowledge, therefore, combines the generative flexibility of LLMs with the factual rigor of structured data, enabling more accurate and trustworthy reasoning over complex domains~\citep{luo2023rog, dong2023hierarchy,bei2025graphs,yasunaga2021qa,luo2024graph}. Evolving development of GraphRAG has progressed along two complementary research trajectories since the seminal work of \citep{graphrag}. The first following approaches have evolved from LightRAG's \citep{lightrag} vector sparsification techniques to more sophisticated graph-aware methods. Subsequent innovations include GNN-RAG and GFM-RAG \citep{mavromatis2024gnn,gfm}, which employ graph neural networks for enhanced node matching, and HippoRAG 1\&2 \citep{hippo,hipporag2} that introduced memory mechanisms and personalized PageRank algorithms for context-aware retrieval. While another Group of methods have focused on improving the quality of knowledge organization, hierarchical approaches like RAPTOR \citep{raptor} and E$^{2}$GraphRAG \citep{e2} employ tree-like clustering and recursive summarization to enhance semantic organization. However, current research remain constrained by their specialized optimizations, either focusing on retrieval or construction in isolation, and lack a unified design. This fragmentation limits their performance on complex reasoning tasks requiring tight integration of knowledge organization and retrieval capabilities, which makes it even harder to adjust the entire framework for generalizability especially when domain shifts occur. Our work bridges this gap by developing a holistic framework that jointly optimizes both aspects while maintaining graph foundation model properties.

\section{Conclusions}
In this paper, we propose \texttt{Youtu-GraphRAG}, a vertically unified agentic paradigm that jointly optimizes both aspects through a graph schema. Our framework introduces \((i)\) a schema-guided agent for continuous knowledge extraction with predefined entity types, relations, and attributes; \((ii)\) dually-perceived community and keyword detection, fusing structural topology with subgraph semantics to construct a hierarchical knowledge tree that supports top-down filtering and bottom-up reasoning; \((iii)\) an agentic retriever interprets the schema to break complex queries into tractable sub-queries, paired with an iterative reasoning and reflection; and \((iv)\) Anonymity Reversion, a novel task to mitigate knowledge leakage in LLMs, deeply measuring the real performance of GraphRAG frameworks supported by a carefully curated anonymous dataset. Extensive experiments across six challenging benchmarks demonstrate \texttt{Youtu-GraphRAG}’s robustness, advancing the Pareto frontier with up to 90.71\% reduction in token costs and 16.62\% higher accuracy than state-of-the-art baselines. Notably, our framework exhibits strong adaptability, enabling seamless domain transfer with minimal schema adjustments. These results underscore the importance of unified graph construction and retrieval, paving the way for more efficient and generalizable GraphRAG.





\setcitestyle{numbers,square}
\bibliography{yoUTU_bib}

\appendix

\section{Prompt templates in LLMs generation}
\label{appendix:A}
\vspace{-1mm}
We present the prompt templates in \ref{pmt:reject} and \ref{pmt:reject}, which designed to evaluate whether permitting LLMs to utilize its parametric knowledge within the RAG system affects performance. To minimize confounding factors, we employed minimalistic prompts that solely differentiate between the two modes.
\subsection{Reject mode}
\label{pmt:reject}
\begin{tcolorbox}[
    colback = blue!5!white, colframe = blue!75!black, arc = 3pt, boxrule = 1pt
]
Given the question and the extracted knowledge from different retrieval paths, please answer the question below.
If the extracted knowledge is not enough to answer, please reject to answer.
\\
\\
Question: \{query\}
\\
\\      
Extracted Knowledge:
\{context\}
\\
\\
Answer:
\end{tcolorbox}
\subsection{Open mode}
\label{pmt:open}
\begin{tcolorbox}[
    colback = blue!5!white, colframe = blue!75!black, arc = 3pt, boxrule = 1pt
]
Given the question and the extracted knowledge from different retrieval paths, please answer the question below. If the extracted knowledge is not enough to answer, please answer it based on your own knowledge.                         
\\
\\
Question: \{query\}
\\
\\      
Extracted Knowledge:
\{context\}
\\
\\
Answer:

\end{tcolorbox}

\section{Data Collection and Processing}
\label{appendix:B}
All raw data in this study are sourced from the original texts of four classic novels: Water Margin, Dream of the Red Chamber, Moby-Dick, and Middlemarch. The copyrights of all these works have entered the public domain, thus presenting no copyright issues. In selecting data sources, we pursued two key objectives: (1) Ensuring comprehensive multilingual evaluation coverage, while
(2) Maintaining sufficient complexity in entity representations (e.g., persons, locations) to rigorously assess model capabilities. The basic statistical information of the dataset is in Table~\ref{tab:question_distribution}.

In our data processing methodology, we employed DeepSeek for entity extraction from the corpus, then the data chunks are anonymized with the extracted entities. Query-answer pairs were constructed by DeepSeek using queries from 2Wiki and MuSiQue as seed templates. Upon acquiring the question-answer pairs, we performed entity anonymization using the same anonymization dictionary as applied to the corpus. This procedure ensures that LLMs cannot effectively leverage parametric memorized patterns from questions. A representative example of anonymized question-answer pairs is presented in Table ~\ref{tab:zero-shot compare}. As clearly demonstrated, while LLMs could handle questions according to common sense knowledge, their performance significantly degrades when confronted with anonymized versions of these questions. This phenomenon forces LLMs to rely on retrieved contextual information rather than depending solely on their parametric knowledge.

To avoid the variance in evaluating subjective questions,  we finally converted the questions into two formats:

\textbf{Anonymity Reversion}. We provide LLMs with anonymized question-answer pairs as context, requiring to infer and reconstruct the original entities that were anonymized. This task specifically assesses the model's ability to leverage contextual clues for entity recovery.

\textbf{Multiple Choice}. To diversify question types and ensure objective evaluation, a subset of questions was converted into multiple-choice format.

We then performed zero-shot filtering to verify model performance on these transformed questions. This design preserves the original assessment objectives of testing the LLM's contextual reasoning capabilities while guaranteeing answer objectivity and uniqueness. Crucially, it mitigates potential unreliability introduced by LLM-as-judge evaluation paradigms. Table ~\ref{tab:final qa} presents representative cases of these two question formats.

\begin{table}[h]
\centering
\caption{Question Type and Difficulty Distribution Statistics}
\label{tab:question_distribution}
\resizebox{\textwidth}{!}{
\footnotesize
\begin{tabular}{@{}c|c|cc|cc|c@{}}
\toprule
\multirow{2}{*}{\textbf{Question Type}} & \multirow{2}{*}{\textbf{Difficulty Level}} & \multicolumn{2}{c}{\textbf{Chinese Literature}} & \multicolumn{2}{c}{\textbf{English Literature}} & \multirow{2}{*}{\textbf{Total}} \\
\cmidrule(lr){3-4} \cmidrule(lr){5-6}
 & & \textit{Water Margin} & \textit{Dream of Red Chamber} & \textit{Moby-Dick} & \textit{Middlemarch} & \\
\midrule
\multirow{2}{*}{\textbf{Objective}} & Simple ($\leq$4 hops) & 29 & 26 & 54 & 54 & 163 \\ 
 & Complex ($>$4 hops) & 24 & 34 & 51 & 22 & 131 \\
 \midrule
\addlinespace[0.2cm]
\multirow{2}{*}
{\textbf{\makecell[c]{Subjective \\ (Anonymous Reversion)}}} 
 & Simple ($\leq$4 hops) & 65 & 50 & 51 & 116 & 282 \\
 & Complex ($>$4 hops) & 146 & 314 & 142 & 219 & 821 \\
\midrule
\multicolumn{2}{c|}{\textbf{Total}} & \textbf{264} & \textbf{424} & \textbf{298} & \textbf{411} & \textbf{1397} \\
\bottomrule
\end{tabular}}
\end{table}

\begin{CJK*}{UTF8}{gbsn} 
\begin{table}[h]
\centering
\caption{Comparations of anonymized and Non-anonymized questions}
\label{tab:zero-shot compare}
\rowcolors{2}{gray!10}{white}
\footnotesize
\resizebox{\textwidth}{!}{
\begin{tabular}{>{\raggedright}p{0.8cm}>{\raggedright}p{0.6cm}p{4cm}p{6.5cm}p{0.8cm}}
\hline
\rowcolor{gray} \textcolor{white}{\textbf{Lang}} & \textcolor{white}{\textbf{Type}} & \multicolumn{1}{>{\centering}p{4cm}}{\textcolor{white}{\textbf{Question}}} & \multicolumn{1}{>{\centering}p{6.5cm}}{\textcolor{white}{\textbf{Zero-shot LLM Response}}} & \textcolor{white}{\textbf{Val}} \\
\midrule
\multirow{2}{*}{CHS} & Raw Query & {从史进拜师到最终离开少华山，期间共经历了哪些身份转变？} & {从富家子弟→王进徒弟→庄主→强盗同伙→落草为寇，共经历四次身份转变。} &  Correct \\ \cmidrule[0.5pt](l){2-5}
& Annoy Query & {从[PERSON\#277]拜师到最终离开[LOCATION\#759]，期间共经历了哪些身份转变？} & {从拜师到离开[LOCATION\#759]，[PERSON\#277]的关键身份转变可梳理如下：1. 入门弟子阶段, 初始以普通学徒身份拜入师门，主要承担基础杂役工作…} & Wrong \\ \midrule
\multirow{2}{*}{ENG} & Raw Query & {What does the narrator's reaction to Queequeg's smoking in bed reveal about their relationship?} & {The narrator's acceptance of Queequeg's smoking reveals their deepening friendship and shared comfort.} &  Correct \\ \cmidrule[0.5pt](l){2-5}
& Annoy Query & What does the [PERSON\#1999]'s reaction to [PERSON\#200]'s smoking in bed reveal about their relationship? & To analyze what [PERSON\#1999]'s reaction to [PERSON\#200]'s smoking in bed reveals about their relationship, we would need more context about the specific reaction and… & Wrong\\ \bottomrule
\end{tabular}}
\label{tab:zero_shot2}
\end{table}
\end{CJK*}

\begin{CJK*}{UTF8}{gbsn} 
\begin{table}[h]
\centering
\caption{Final Question-Answer Formats}
\label{tab:final qa}
\footnotesize
\resizebox{\textwidth}{!}{
\begin{tabular}{>{\raggedright}p{0.6cm}>{\raggedright}p{8cm}p{4cm}}
\hline
\rowcolor{gray} \textcolor{white}{\textbf{Lang}} &
\multicolumn{1}{>{\centering}p{8cm}}{\textcolor{white}{\textbf{Question}}} & \multicolumn{1}{>{\centering}p{4cm}}{\textcolor{white}{\textbf{Ground Truth}}} \\
\midrule
\multicolumn{3}{c}{\textbf{Anonymity Reversion}}\\
\midrule
CHS & \makecell[l]{请根据上下文对下面这段问答\\
\symbol{96}\symbol{96}\symbol{96} \\
Q: 在[PERSON\#532]离开[LOCATION\#526]后，他在哪个村庄\\的酒店中与[PERSON\#277]重逢？这个村庄附近的山
上盘踞着\\哪两位头领？\\
A: [PERSON\#532]在[LOCATION\#110]附近的酒店与[PERSON\\\#277]重逢，该村庄附近的[LOCATION\#535]上盘踞[PERSON\#\\503]和[PERSON\#4]两位头领。\\
\symbol{96}\symbol{96}\symbol{96} \\
\\
中已经被匿名化处理的所有人名和地名等进行推理，判断
出被\\匿名的原本内容是哪些。} & \makecell[l]{PERSON\#532——鲁智深\\PERSON\#277——史进\\PERSON
\#4——周通\\PERSON\#503——李忠\\LOCATION\#526——五台山\\LOCATION\#110——桃花村\\LOCATION\#535——桃花山}  \\ 
\midrule
ENG & \makecell[l]{Please read the following QA pairs\\
\symbol{96}\symbol{96}\symbol{96} \\
Q: What does [PERSON\#200]'s story about the wedding feast\\ reveal about cultural misunderstandings?\\
A: The story reveals how cultural misunderstandings, such as \\  
{[PERSON\#588]} mistaking the punchbowl for a finger-glass, can \\ arise from ignorance of local customs.\\
\symbol{96}\symbol{96}\symbol{96} \\
\\
then for all anonymized Persons and Locations, perform infer-\\ence to determine the original content that was anonymized.} &
\makecell[l]{PERSON\#200——Queequeg\\PERSON\#588——captai}
\\ \midrule
\multicolumn{3}{c}{\textbf{Multiple Choice}}\\
\midrule
CHS & \makecell[l]{海棠诗社成立时，[PERSON\#315]给自己取的别号是什么？这个\\别号与她居住的哪个场所相关？\\ \\
A. [LOCATION\#340]；[LOCATION\#625]老农\\
B. [LOCATION\#340]；[LOCATION\#340]隐士\\
C. [LOCATION\#625]老农；[LOCATION\#340]\\
D. [LOCATION\#340]居士；[LOCATION\#625]老农}&
\makecell[l]{C. (李纨，稻香老农，稻香村)}\\
\midrule
ENG & \makecell[l]{Which two physical traits do [PERSON\#1035] and her daughter \\ {[PERSON\#445]} share in common?\\\\
A. Straight hair and round faces \\
B. Curly hair and square faces \\
C. Wavy hair and oval faces \\
D. Short hair and triangular faces} &
\makecell[l]{B. (Mrs. Garth, daughter Mary)}\\

\bottomrule
\end{tabular}}
\label{tab:zero_shot1}
\end{table}
\end{CJK*}






\end{document}